\documentclass[twoside,twocolumn,fleqn,superscriptaddress,showkeys]{revtex4-2}
\usepackage{xcolor}    
\usepackage{graphicx}  
\usepackage{amssymb}   
\usepackage{amsmath}   
\usepackage{bm}        
\usepackage{times}     
\usepackage{fancyhdr}  
\begin{document} 

\title{A Spin model for global flat-foldability of random origami} 
\author{Chihiro Nakajima} 
\affiliation{Tohoku Bunka Gakuen University, Sendai Miyagi 6-45-1, Japan} 
\email{chihiro.nakajima@ait.tbgu.ac.jp}

\date{\today} 

\begin{abstract}
We map the problem of determining flat-foldability of the origami diagram onto the ground-state search problem of spin glass model on random graphs. If the origami diagram is locally flat-foldable around each vertex, a pre-folded diagram, showing the planar-positional relationship of the facet, can be obtained. For remaining combinatorial problem on layer ordering of facets can be described as a spin model. A spin variable is assigned for the layer-ordering of each pair of facets which have an overlap in the pre-folded diagram. The interactions to prohibit the intrusion of each facet into the other component of the same origami diagram are introduced among two or four spins. The flat-foldability of the diagram is closely related to the (non-)existence of frustrated loops on the spin model with the interactions on the random (hyper)graph.
\end{abstract}

\keywords{Computational origami, Statistical mechanics, Flat-foldability, Random spin model}

\maketitle  
\thispagestyle{fancy}  

\section{INTRODUCTION}
Origami has its origin in traditional artwork of folding a piece of paper along straight crease lines into various shapes of animals, plants, or objects.
By the extensive research in various fields, origami has found a number of interesting mathematical and engineering applications for today.
In order to fold a sheet along straight crease lines into many kinds of geometric patterns, conformation of facets and creases should be incorporated into a sheet of paper in a well-planned manner.
This sheet with planned facets and creases is called an origami diagram or a crease pattern.
The design of the origami diagram has been one of research subjects in the field of computer seience.

The problem to determine whether a given origami diagram is actually foldable into a flat plane, is called the flat-foldability problem, which is one of the major topics in the computational origami\cite{Hull_t}. 
On the computational complexity class, this problem is proven to be NP-hard\cite{BH} and 
it is known that to assign the global layer ordering of flaps, namely closed polygonal facets separated by creases, can be the central source of its hardness\cite{BH,Assis}. 

Analyses of foldability properties associated with spin models have been previously performed in somewhat limited ways, each with interesting results\cite{GH_M,Assis}.
Furthermore, there are also works dealing with the folding of the membrane, that is a physical reality of folding itself under the thermal fluctuation, with the treatment by spin models\cite{FG,XM}.
Previous statistical mechanics studies on origami have focused on the proximity of facets on origami diagrams, and have dealt with this by relating to the proximity of interactions on models.
While, we consider the overlap for every pair of facets, including pairs for not close, or connected, on the origami diagram. 
We pay attention to whether two facets have the overlapped region while illustrating the positional relationship between two facets after the origami diagram is folded along the creases.
With such manner we can obtain a complete correspondence with the model of random spin systems.
In fact, it was previously pointed out that the global layer ordering of facets should in principle be translated into a statistical mechanics model\cite{Assis}.
However, by actually doing this, it is possible to highlight things that are not apparent from the proximity on the origami diagram.
We can find the elements that give unfoldability in a more intuitive way on a hypergraph composed of interactions.
In addition we can expect a reduction in the amount of computation in the flat-foldability problem by considering the structure of the hypergraph.

\section{Models and Implementations}
\begin{figure}
\includegraphics[width=8.5cm]{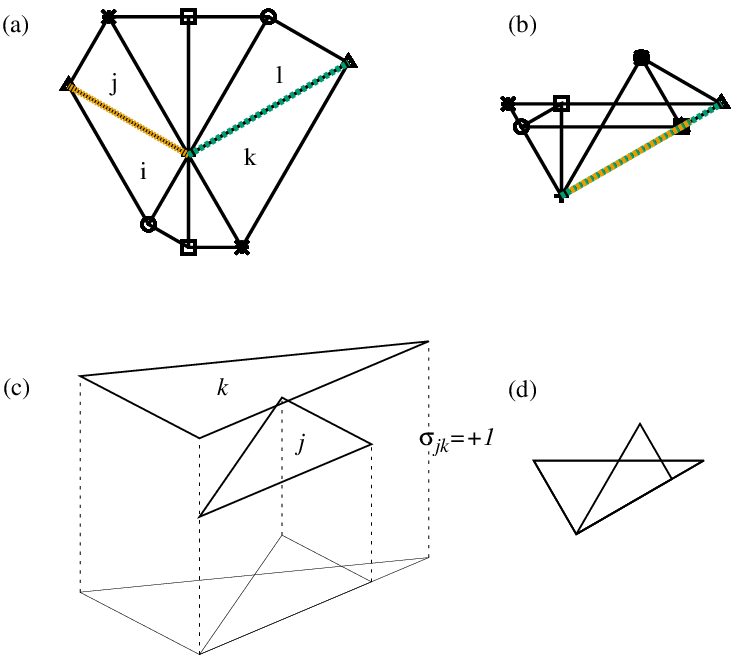} 
\caption{(a)Example of origami diagram. Each edge in the figure represents a crease. In this figure there are no overlaps of facets. (b)Corresponding pre-folded diagram, which describes the overlaps of facets when the figure (a) is folded along the creases. 
Each mark corresponds to the vertex in Fig.(a) with the same mark. Unmarked cross-points in Fig.(b) are simply the intersection of two edges caused by the overlaps, not the vertex of a facet.
(c)Schematic picture of introduction of the Ising variable to a local layer-ordering between the facets $j$ and $k$. (d)The planar-positional relationship of the facets $j$ and $k$ corresponding to Fig.(c), also included in Fig.(b).
}\label{fig:fig1}
\end{figure}
There are Kawasaki's theorem which gives necessary and sufficient conditions for the (local) flat-foldability of the origami diagrams which consist of one interior vertex and other surrounding exterior vertices\cite{Kaw_s}.
According to the theorem, the number of facets clustered around the central vertex must be even and the alternating sum of the angles clustered around the vertex must be $0$.
For an origami diagram that satisfies the conditions from Kawasaki's theorem for every internal vertices, it is obtained a second diagram describing the planar-positional relationship of the facets after folding every creases.
Here this second diagram is called pre-folded diagram.
For example, Fig.\ref{fig:fig1}(a) is an origami diagram which satisfies the condition with Kawasaki's theorem.
The eight polygons (triangles) included in this figure are all facets, and the edges connecting the facets are creases.
When the eight creases in Fig.\ref{fig:fig1}(a) are folded, the shape becomes as Fig.\ref{fig:fig1}(b).
Folding causes the overlaps among facets.
This Fig.\ref{fig:fig1}(b) is the pre-folded diagram corresponding to the origmai diagram of \ref{fig:fig1}(a).
Hence, the (global) flat-foldability problem is as a problem of determining whether or not each facet can be stacked without breaking, that is, the intrusion of a facet of the origami paper into other component in the same paper.

Next, according to the pre-folded diagram, the layer-ordering of the facets should be determined.
The local layer ordering, whether one facet of the pair is below- or above- side of the another, can be represented by the binary variable.
Therefore the Ising spin variables $\sigma \in \{+1,-1\}$ are assigned for each pair of facets which have an overlap in the pre-folded diagram.
For example, in the case of the figure in Fig.\ref{fig:fig1}(c) $\sigma_{j,k}$ is defined to take its value $+1$ when the facet $k$ is above the facet $j$ in $z$-axis (the direction of height).
Here, $\sigma_{j,k}$ is also defined to invert its value for the exchange of the order of indices, namely $\sigma_{j,k}=-\sigma_{k,j}$.
From the set of $\{\sigma\}$ which describes the local layer orderings, the total layer ordering of facets is reconstructed.
The variables $\sigma_{j,k}$ according to the above definition are convenient to give terms of the energy function described later, as Eqs.(\ref{eq:term_J}), (\ref{eq:terms_K}), and (\ref{eq:terms_L}), using spin variables from the geometry of facets via subscripts.
However, from the point of view of statistical mechanics, it obscures the prospects for unambiguously defining the state space or the configuration space to use the variables with exchange of the order of subscripts or sign-reversal.
To ensure better prospect, we introduce the variables $s_{j,k}$ and $\tau_{jk}$ as follows and use them to rewrite $\sigma_{j,k}$.
\begin{eqnarray}\label{eq:penetation_2}
\tau_{jk}&=&\mathrm{sign}(k-j),\\
s_{j,k}&=&\sigma_{\mathrm{min}(j,k),\mathrm{max}(j,k)}.
\end{eqnarray}
With the variables $\sigma_{j,k}$, $s_{j,k}$ and $\tau_{jk}$, the following relation holds,
\begin{eqnarray}\label{eq:penetation}
\sigma_{j,k}&=&\tau_{jk}s_{j,k}.
\end{eqnarray}
The behavior of the variable for the exchange of the order of subscripts is, respectively,
\begin{eqnarray}
\sigma_{j,k}=-\sigma_{k,j}, \label{eq:rule_sigma}\\
\tau_{jk}=-\tau_{kj}, \label{eq:rule_tau}\\
s_{j,k}=s_{k,j} \label{eq:rule_s}.
\end{eqnarray}
The variable $s_{j,k}$ also trivially takes two values $+1$ or $-1$.
By considering the state space in terms of $s_{j,k}$, the state sum is taken on the state space of $\{s_{j,k}\}$, where $j<k$.
Thus the bounds of the state space for the origami-diagram with $N$ facets are limited to $2^{N(N-1)/2}$.
After here, when the spin variable $s_{a,b}$ in which the former part of whose subscript is larger than the latter one appears, it shall be received according to Eq.(\ref{eq:rule_s}).
One might be concerned that different styles of facet labeling will have different signs for the above variables even for the same pre-folded diagram.
In fact, however, we can label the facets in any order, for reasons based on gauge transformations described afterward.

By representing the problem with the combination of the local layer-ordering, the flat-foldability problem of the origami becomes the combinatorial optimization problem.
Under the planar placement of each facet in the pre-folded diagram, depending on the choice of total layer-ordering, there are some cases in which an intrusion between facets may become unavoidable.
To prohibit such ordering, some cost function terms are introduced.
Constrains are introduced so that the energy takes a positive value for prohibited configurations.
For the spin configurations that satisfy all constraints, they will be corresponding to the realizable layer-ordering.

As the first kind of constraints, we introduce a term that contains a product of two spin variables to prohibit an intru- sion of a facet into a crease. 
If one facet $k$ has an overlap with other two facets $i,j$, connected to each other, in the pre-folded diagram, the former may be sandwiched between the latter so that it intrude into the junction of the two facets.
Such an arrangement is not realizable. 
The case which the facet $k$ is sandwiched between $i$ and $j$ is corresponding to combinations of spin that is $\{ \sigma_{i,k},\sigma_{k,j}\}=\{1,1\}$ or $\{-1,-1\}$.
\begin{eqnarray}
E^{(i)}_{ij,k}&=&\frac{1}{2}\big(1+\sigma_{i,k}\sigma_{k,j}\big)\nonumber\\
&=&\frac{1}{2}\big(1-J_{(ik)(kj)}s_{i,k}s_{k,j}\big)\qquad\qquad\qquad\qquad \label{eq:term_J}\\
J_{(ik)(kj)}&=&-\tau_{ik}\tau_{kj}.\label{eq:g_inv_J}
\end{eqnarray}

\begin{figure}
\includegraphics[width=6cm]{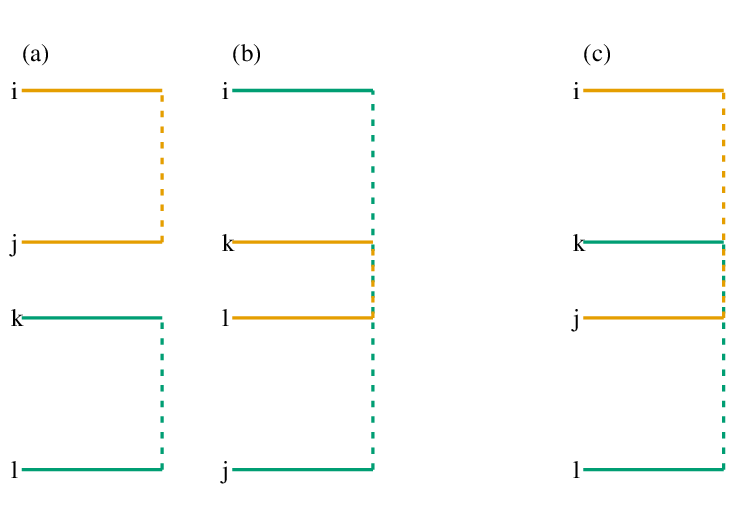} 
\caption{(a),(b)Layer orderings of facets which are accepted under the coincident of two creases.
(c) Layer ordering which is NOT accepted under the coincident of two creases. One connected pair of facets must penetrate the another connected part of the remanent two facets.
}
\label{fig:fig2}
\end{figure}

The second kind of terms are required if the two creases are in geometrically coincidental position in the pre-folded diagram.
Suppose that there are two creases that partially coincide with each other and pairs of facets, called $i,j$ and $k,l$, each connected by the creases.
The example is exhibited in Fig.\ref{fig:fig1}(a) and (b).
For such arrangements and connections of facets there are no intrusion even if both $i$ and $j$ are sandwiched between $k$ and $l$, different from the cases considered in Eq.(\ref{eq:term_J}) and (\ref{eq:g_inv_J}).
However, if only one of $i$ or $j$ is sandwiched between $k$ and $l$, the creases will penetrate each other at the region where two connections coexist.
In Fig.\ref{fig:fig2}(a)-(c), the schematic picture of the local layer-ordering of the four facets $i$, $j$, $k$ and $l$ in the origami diagram and the pre-folded diagram of Fig\ref{fig:fig1}(a) and (b), an example of two pairs of facets connected by creases with partial coincidence in planar-positioning.
The dotted lines of color orange and green in Fig.\ref{fig:fig2} are with respect corresponding to the edges connecting $i$ and $j$, $k$ and $l$ in Fig.\ref{fig:fig1}(b).
The each pair of facets connected to the both side of each edge, $i$, $j$, $k$ and $l$, is represented as solid line with the same color of the edge.
Fig.\ref{fig:fig2}(a) and (b) represent the acceptable layer-orderings.
Especially Fig.\ref{fig:fig2}(b) is corresponding to the cases that both $k$ and $l$ are sandwiched between $i$ and $j$.
While, Fig.\ref{fig:fig2}(c) is corresponding to the prohibited layer-ordering that only $k$ is samdwitched between $i$ and $j$.
In these cases we need the product of four spin variables as the following form, 
\begin{eqnarray}
E^{(q)}_{ijkl}&=&\frac{1}{2}\Big(1-\sigma_{i,k}\sigma_{i,l}\sigma_{j,k}\sigma_{j,l}\Big) \nonumber\\
&=&\frac{1}{2}\Big(1-K_{ijkl}s_{i,k}s_{i,l}s_{j,k}s_{j,l}\Big),\label{eq:terms_K}\\
&&K_{ijkl}=\tau_{ik}\tau_{il}\tau_{jk}\tau_{jl}.\label{eq:g_inv_K}
\end{eqnarray}

In addition to the above two kinds, the third kind of constraints is imposed to prohibit cyclic ordering of three facets.
For three facets $i$, $j$, and $k$ with $i<j<k$, cyclic ordering, for example the case represented with $\{\sigma_{i,j},\sigma_{i,k},\sigma_{i,k}\}=\{1,1,-1\}$, can be considered in general.
However, in the case these three facets have areas shared with each other in the pre-folded diagram, such ordering is unrealizable.
To prohibit the cyclic ordering, the set of interaction terms of spins are introduced as follows,
\begin{eqnarray}
E^{(c)}_{ijk}&=&\frac{1}{4}\big(1+\sigma_{i,j}\sigma_{j,k}+\sigma_{j,k}\sigma_{k,i}+\sigma_{k,i}\sigma_{i,j}\big) \label{eq:prohibitcycle}\nonumber\\
&=&\frac{1}{4}\big(1-L_{(ij)(jk)}s_{i,j}s_{j,k}\nonumber\\
&&\!\!\!\!-L_{(jk)(ki)}s_{j,k}s_{k,i}-L_{(ki)(ij)}s_{k,i}s_{i,j}\big), \label{eq:terms_L}\\
L_{(ij)(jk)}&=&-\tau_{ij}\tau_{jk}
\end{eqnarray}
Notice that
there are such cases for 3 facets that no regions where all of them stack at the same time while any 2 pairs of them have each overlap.
For these cases the set of terms like Eq.(\ref{eq:prohibitcycle}) is not to be introduced because the cyclic ordering is actually realizable without intrusion.

In the process of labeling indices into facets on an pre-folded diagram, any order can be employed.
The sign of each interaction coefficient $J_{(ij)(jk)}$, $K_{ijkl}$, and $L_{(ij)(jk)}$ varies depending on the details of the index-labeling of facets.
However, the changes of them due to the differences of the way of labeling can be rewritten back to the original Hamiltonians by the following transformations with gauge variables $u_{ij} \in \{1,-1\}$ such as
\begin{eqnarray}\label{eq:gauge_tf_2}
\!\!\!\!\!\!\!\!\!\!\! J_{(i'k')(k'j')} &\rightarrow& \hat{J}_{('i'k)(k'j')}=u_{i'k'}u_{k'j'}J_{(i'k')(k'j')},\\
\!\!\!\!\!\!\!\!\!\!\! K_{i'j'k'l'} &\rightarrow& \hat{K}_{i'j'k'l'}'=u_{ik}u_{i'l'}u_{j'k'}u_{j'l'}K_{i'j'k'l'},\\
\!\!\!\!\!\!\!\!\!\!\! L_{(i'k')(k'j')} &\rightarrow& \hat{L}_{(i'k')(k'j')}=u_{i'k'}u_{k'j'}L_{(i'k')(k'j')},
\end{eqnarray}
where the value of $u_{ij}$ is given by the relation between the two different lookings $s_{i,j}$ and $s_{i',j'}$ which are caused by different ways of index-labeling of facets,
\begin{eqnarray}\label{eq:gauge_tf}
u_{i'j'}s_{i',j'}&=&s_{i,j}.
\end{eqnarray}
Since the partition function is invariant under the gauge transformations, the same layer-ordering can be computed with any index-labeling of facets\cite{Nis_p,Nis_t}.

As an example of the above transformation, we see the origami-diagram in Fig.\ref{fig:fig3}.
In the origami-diagram in Fig.\ref{fig:fig3}(a) the indices $i \in \{1,2,3,4\}$ is allocated to each facet.
Suppose that $i' = 4,3,1,2$ is assigned to the each same facet (with $i=1,2,3,4$) by different way of indexing.
A term which prohibits the intrusion of the facet $i=4$ between the two facets $i=2 \quad \mathrm{and} \quad 3$, whose form is $(1-J_{(24)(43)}s_{2,4}s_{3,4})/2$ with $J_{(24)(43)}=1$, is included in the terms in the energy function.
Within the way of indexing with $i' = 4,3,1,2$, the same prohibition term is described as the form $(1-J_{(32)(21)}s_{2,3}s_{1,2})/2$ with $J_{(32)(21)}=-1$.
Note that in these two descriptions $s_{2,4}=1$ within $i=1,2,3,4$ and $s_{2,3}=-1$ within $i'=4,3,1,2$ represent the same local layer-ordering in the sense of geometry.
Thus, the signs of the interaction coefficients differ each other depending on the way of indexing, even though the same in the meaning of the geometric content is prohibited.
However, when we apply the above transformation with the gauge variable $u_{ij}$ as
\begin{eqnarray}
u_{14}=-1\label{eq:gv_exam_1},\\
u_{34}=-1,\\
u_{24}=-1,\\
u_{13}=-1,\\
u_{23}=-1,\\
u_{12}=1\label{eq:gv_exam_z},
\end{eqnarray}
the interaction coefficients are transformed into,
\begin{eqnarray}
\hat{J}_{(32)(21)}=(+1)(-1)J_{(32)(21)}=-J_{(32)(21)},
\end{eqnarray}
and the term $(1-\hat{J}_{(32)(21)}s_{2,3}s_{1,2})/2$ within $i'=4,3,1,2$ becomes the same as $(1-J_{(24)(43)}s_{2,4}s_{3,4})/2$ within $i=1,2,3,4$.
Looking at the same thing as this rewrite for the whole intrusion-prohibiting terms,
at first the terms with $i=1,2,3,4$ is described as,
\begin{eqnarray}\label{eq:phtm_cent}
\sum E^{(i)}_{(ij,k)}&=&\frac{1}{2}\Big(6-s_{1,3}s_{1,2}-s_{1,4}s_{1,2}-s_{1,3}s_{2,3}\nonumber\\
&&\quad\!+s_{3,4}s_{2,3}-s_{1,4}s_{2,4}-s_{3,4}s_{2,4}\Big).
\end{eqnarray}
Similarly, the same contents within $i'=4,3,1,2$ is described as,
\begin{eqnarray}\label{eq:phtm_dash}
\sum E^{(i)}_{(i'j',k')}&=&\frac{1}{2}\Big(6-s_{1,4}s_{3,4}-s_{2,4}s_{3,4}-s_{1,4}s_{1,3}\nonumber\\
&&-s_{1,2}s_{1,3}-s_{2,4}s_{2,3}+s_{1,2}s_{2,3}\Big),
\end{eqnarray}
where the order in which the terms are arranged is the order of the same geometric content, namely the prohibition for same subset of facets, as above.
By the transformation with the gauge variables (\ref{eq:gv_exam_1})-(\ref{eq:gv_exam_z})the form of the prohibition term (\ref{eq:phtm_dash}) is written back to the same as Eq.(\ref{eq:phtm_cent}), as exhibited that
\begin{eqnarray}
\sum \hat{E}^{(i)}_{(i'j',k')}&=&\frac{1}{2}\Big(6-s_{1,4}s_{3,4}-s_{2,4}s_{3,4}-s_{1,4}s_{1,3}\nonumber\\
&&+s_{1,2}s_{1,3}-s_{2,4}s_{2,3}-s_{1,2}s_{2,3}\Big).
\end{eqnarray}

\section{Details of Energy Function for Some Illustrative Examples}
\begin{figure}
\includegraphics[width=8.5cm]{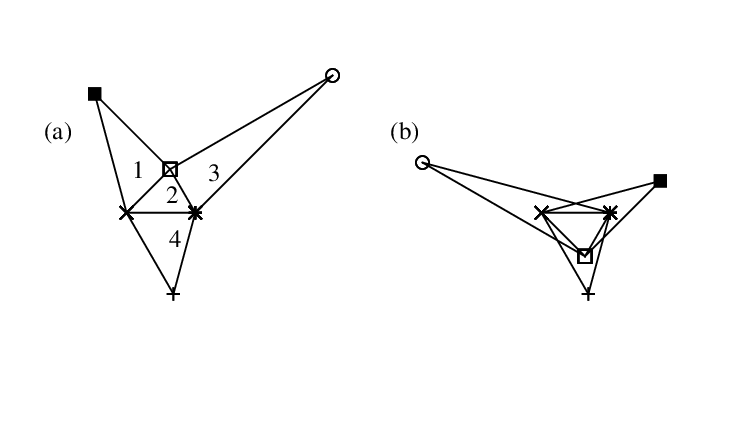} 
\caption{(a)An example of the origami diagram which is flat-unfoldable.
The numbers in the figure represent the indices of the facets. The pairs of numbers appearing in the subscripts of Eqs. (\ref{eq:E_FRUST4_i}) and (\ref{eq:E_FRUST4_c}) and (\ref{eq:frust_gutai_min}) in the text are given corresponding to them.
(b)Pre-folded diagram corresponding to the Fig.(a). Each mark corresponds to the vertex in Fig.(a) with the same mark. 
Unmarked cross-points in Fig.(b) are simply the intersection of two edges caused by the overlaps, not the vertex of a facet.
}\label{fig:fig3}
\end{figure}

In the example of Fig.\ref{fig:fig3}, each facet located outside, $1,3,4$, cannot be sandwiched between every creases which consist of other two facets.
Therefore in the energy function for this pre-folded diagram 6 terms of first kind are included.
In addition there are 4 terms of third kind.
Totally, the form of the energy function for this diagram is as follows,
\begin{eqnarray}
\!\!\!\!\!H\big(\large\{s\large\}\big)&=&\sum E^{(i)}_{(ij,k)} + \sum E^{(c)}_{(ijk)}\\
\!\!\!\!\!\sum E^{(i)}_{(ij,k)}&=&\frac{1}{2}\Big(6-s_{1,3}s_{1,2}-s_{1,4}s_{1,2}-s_{1,3}s_{2,3}\nonumber\\
&&\quad\quad+s_{3,4}s_{2,3}-s_{1,4}s_{2,4}-s_{3,4}s_{2,4}\Big)\label{eq:E_FRUST4_i}\\
\!\!\!\!\!\sum E^{(c)}_{(ijk)}&=&\frac{1}{4}\Big(1+s_{1,2}s_{2,3}-s_{2,3}s_{1,3}-s_{1,3}s_{1,2}\nonumber\\
&&\quad\;+1+s_{1,2}s_{2,4}-s_{2,4}s_{1,4}-s_{1,4}s_{1,2}\nonumber\\
&&\quad\;+1+s_{1,3}s_{3,4}-s_{1,4}s_{3,4}-s_{1,3}s_{1,4}\nonumber\\
&&+1+s_{2,3}s_{3,4}-s_{3,4}s_{2,4}-s_{2,4}s_{2,3}\Big), \label{eq:E_FRUST4_c}
\end{eqnarray}
where the interaction coefficients are expressed numerically.
The terms in Eq.(\ref{eq:E_FRUST4_i}) come from the prohibition for intrusive layer-orderings. These form a cycle.

Similarly, the terms in Eq.(\ref{eq:E_FRUST4_c}) are given from the prohibition of cyclic orderings. 

In this energy function, the chain of 2-spin interactions in the first 2 rows, given from $\{J_{(ij)(jk)}\}$, forms a closed cycle in which the product of $J_{(ij)(jk)}$s involved is negative, that is,
\begin{eqnarray}
\prod_{(ik)(kj)}^{\mathrm{cycle}} J_{(ij)(kl)}=\qquad\qquad\qquad&&\nonumber\\
=J_{(31)(12)}J_{(41)(12)}J_{(13)(32)}&J&_{(43)(32)}J_{(14)(42)}J_{(34)(42)}\nonumber\\
=-1. \qquad\qquad\qquad&&\label{eq:frust_gutai_min}
\end{eqnarray}
In the cycle the $6$ spin variables are involved and then there are $64$ possible binary combinations.
When the sign of the product of $J_{(ij)(jk)}$s involved is negative, none of the combinations can give the value $\sum E^{(i)}=0$ of the local energy corresponding to this cycle.

\begin{figure}
\includegraphics[width=8.5cm]{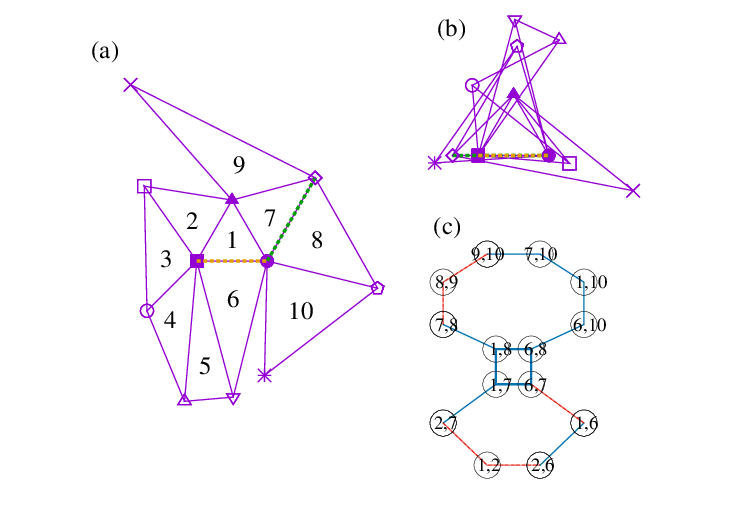} 
\caption{
(a)Another example of the origami diagram which is flat-unfoldable. 
(b)Pre-folded diagram corresponding to the Fig.(a). Each mark corresponds to the vertex in Fig.(a) with the same mark. 
Unmarked cross-points in Fig.(b) are simply the intersection of two edges caused by the overlaps, not the vertex of a facet.
(c)Subgraph with $J_{(ij,k)}$ and $K_{ijkl}$, which is frustrated. 
The numbers $i,j$ inside the circles in the figure represent the subscripts of two facets whose layer-orderings are described by the spin variables $s_{i,j}$. For example, numbers $9,10$ in upper-left circle represents $s_{9,10}$.
The solid blue line represents the ferromagnetic interaction, and the dashed red line represents the antiferromagnetic interaction.
}\label{fig:fig4}
\end{figure}

When 4-body interactions are involved, frustration is defined not for a single cycle, but for a composite structure of multiple cycles combined through 4-body interactions.
In the example of the origami and pre-folded diagram in Fig.\ref{fig:fig4}, the two creases each connecting facets 1, 6 and 7, 8 in (a) are coincident in (b).
Hence, the Hamiltonian corresponding to this pre-folded diagram contains four-body interactions involving facets 1, 6, 7, and 8, whose interaction coefficient is represented as $K_{1678}$.
In the Hamiltonian, totally there are 42 terms with the form same as Eq.(\ref{eq:term_J}), 84 terms with the form same as Eq.(\ref{eq:terms_L}), and a term with the form same as Eq.(\ref{eq:terms_K}).
Then, 11 terms in $\sum E^{(i)}+\sum E^{(q)}$ form a composite of 2 cycles in which the sign of the product of interaction coefficients involved is negative as follows,
\begin{eqnarray}
&\prod_{(ij)(jk)}^{\mathrm{cycle}}&J_{(ij)(jk)}\prod_{(mnop)}^{\mathrm{cycle}}K_{mnop}\nonumber\\
&&\!\!\!\!\!\!\!\!\!\!\!\!\!\!\!\!\!\!\!\!\!\!\!\! =J_{(1,7)(7,2)}J_{(7,2)(2,1)}J_{(1,2)(2,6)}J_{(2,6)(6,1)}J_{(1,6)(6,7)}\nonumber\\ 
&&\!\!\!\!\!\!\!\!\!\!\!\!\!\!\!\!\!\!\!\!\!\!\!\! \times J_{(1,8)(8,7)}J_{(7,8)(8,9)}J_{(8,9)(9,10)}J_{(9,10)(10,7)}\nonumber\\
&&\!\!\!\!\!\!\!\!\!\!\!\!\!\!\!\!\!\!\!\!\!\!\!\! \times J_{(7,10)(10,1)}J_{(1,10)(10,6)}J_{(10,6)(6,8)}\nonumber\\
&&\!\!\!\!\!\!\!\!\!\!\!\!\!\!\!\!\!\!\!\!\!\!\!\! \times K_{1678}\nonumber\\
&=&-1. \label{eq:frust_gutai_quad}
\end{eqnarray}
This diagram is also unfoldable.

In a model in which all interaction coefficients are given only by binary variables $+1$ or $-1$, it generally holds that there is no spin configurations that satisfy all the 2- (or 4-) spin interactions in the cases which the subgraph forms a cycle (or a composite of them) and the product of the signs of the interaction coefficients $J_{(ij)(jk)}$ (or $K_{ijkl}$) involved is negative.
Therefore, if the model graph contains frustration, for example as illustrated here, we can conclude that the configurations with $\sum E^{(i)}_{ij,k} + \sum E^{(q)}_{ijkl}=0$ does not exist.
Same as the partition function, the product of interaction coefficients $J_{(ij,k)}$ or $K_{ijkl}$ contained in the closed cycle or its complex, a quantity called frustration, is also invariant to the gauge transformation shown in Eq.(\ref{eq:gauge_tf}).
If we rewrite the expressions of Eq.(\ref{eq:frust_gutai_min}) or Eq.(\ref{eq:frust_gutai_quad}) with $\tau_{jk}$s according to Eq.(\ref{eq:g_inv_J}) and (\ref{eq:g_inv_K}), we see that each $\tau_{jk}$ corresponding to the same pair of the facet-indices always appears an even number of times.
Therefore changes in the values of $J_{(ij,k)}$s and $K_{ijkl}$s accompanying the changes in the labeling order of facets are canceled in the cycle.
Hence, the detection of the presence of frustration can be used to determine unfoldability.

This section focuses on frustration caused by the intrusion prohibition. 
In the case that the frustration caused by them is included in the Hamiltonian, no ground states with $H=0$ exist.
Hence, 
the searching for such frustration is certainly a part of discriminating the unfoldability.
The cyclic prohibition terms do not prohibit any total layer ordering that does not include cyclic ordering.
In this sense, it can be said that there is no non-trivial prohibition caused by cyclic prohibition terms.
While, it is possible situation that only the layer orderings which contain cyclic order are allowed, namely be the ground states, in the Hamiltonians without frustration.
Such situations are, in present, in need of further consideration.

\section{Further Consideration on the Hypergraph Obtained from Spin Model}
Hypergraphs formed by interactions obtained by rewriting to the spin model often contain branches extending from the core.
The spins included in these branches can take values depending on those on the root side adjacent, as long as the branches have open extremities.
Since the branches does not contribute to the frustration, we are allowed to consider the hypergraph with their branches removed, once after given according to the procedure in the previous section, for the purpose of detection of the frustration.

In addition, performing this removal in combination with the connected component analysis leads further reduction of computational effort.
This is because there are cases that some arbitrary of folding way appear as a connected component which is independent, at least with respect to $J_{(ij,k)}$ and $K_{ijkl}$, from the other ones in the hypergraph. 
Both removal of the branches and connected component analysis can be done in polynomial time for the system size, \textit{i.e.} the number of spins.

The problem of detecting frustration in partial hypergraphs still seems to face a combinatorial optimization problem. So it is true that the approaches presented in the present and the previous section may not provide significant speedups to the problem, such as reducing the exponential dependence to the polynomial dependence on system size. However, reducing the effective system size to the number of spins which is involved in the core of the largest connected component also multiplicatively reduces the amount of computation.
Therefore we can realistically expect to reduce the amount of computation by rewriting the problem of detecting frustration in the spin model.

By using methods such as the Monte Carlo method and a thermodynamic integration or a reweighted histogram method, it is possible to detect the unfoldability and evaluate the total number of layering orders at the same time through the evaluation the number of states.
This method allows the analysis of the energy landscape based on the spin model for each single instance of origami diagram.
Although the Monte Carlo methods do not necessarily mean speeding up the algorithm, they are undoubtedly important as a realistic and highly versatile numerical method.

\section{CONCLUSIONS}
In summary, in this proceeding we show that by translating the origami diagram into a spin model
it is possible to describe the unfoldability criterion in a natural way, as frustration, 
and also expectable to reduce the computational effort with focusing on the cycles and connected components of the hypergraph obtained from the model.

Formulating the problem by spin model is the first step to apply various statistical mechanics methods for random systems, including the replica method, to the flat-foldability determination problem.
One of the advantages of the statistical-mechanical treatment is extracting information about the average-case properties and its parameter dependence through the analysis with replica method.
It is expected that the whole picture of the flat-foldability problem of origami will be further clarified, for example, the existence or the location of the phase transition which divides the parameter-region into one whose most instances are flat-foldable and one whose most instances are unfoldable, by examining the average properties for sets of many instances.
It is expected that the whole picture of the flat-foldability problem of origami will be further clarified by examining the average-case properties for sets of many instances.

Among them, the average-case computational complexity is an important topic, which is expected to be obtained through the analysis of replica symmetry breaking\cite{MZKST,WH,KMRSZ}.
In the sense of the computational complexity class, which is given based on the worst-case complexity, the problem dealt with in this research has already proven with the polynomial reduction to be NP-hard, as mentioned in the introduction. However, the analysis of the average-case computational complexity has not progressed relatively to that of the worst-case ones.

However, in order to apply the replica method it is necessary to give some probabilistic model for the distribution or the probability of the realization of each instance, that is, each origami diagram in this problem.
Probabilistic generation of locally flat-foldable origami diagrams is a non-trivial problem, and this problem must first be overcome in order to carry out the replica calculation.
Alternatively, if we can build the model that is a randomized version of the origami diagram which gives an upper bound on the free energy, or mathematically find a polynomial reduction from the random spin model to the origami diagram, they may provide the gateway which leads to the replica analysis.

The spin-model formulation explained in this proceeding has also some direct extendability to other computational problems related to origami.
For example, there is a map folding problem; the folding of the region tessellated into uniformly rectangular facets\cite{ABDDBSSandM}.
Variants of this problem range from those known to be solved with polynomial-time-algorithms to those known to be in the class NP- complete. 
In the map folding problem, including the variants, the manner of folding, mountain or valley, is assigned for each crease.
This problem can be translated with the procedure explained in this proceeding into the Ising-spin model consisting of 4-body interactions and a random field for each spin.
For some of the problems in the range of the extension, including the above example, it is expected to be relatively easy to obtain the average-case properties.


\begin{references} 
\bibitem{Hull_t}T. C. Hull, \textit{Origametry:Mathematical Methods in Paper Folding} (Cambridge Univ. Press, Cambridge, 2020), T. C. Hull, \textit{The combinatorics of flat folds: a survey}, in \textit{Origami3} (Natick, MA: A. K. Peters, 2001), pp. 29–38. 
\bibitem{BH}M. Bern and B. Hayes, in \textit{Proceedings of the Seventh Annual ACM-SIAM Symposium on Discrete Algorithms}, (Society for Industrial and Applied Mathematics, Philadelphia, 1996), p. 175-183.
\bibitem{GH_M}J. Ginepro and T. C. Hull, J. Int. Seq. \textbf{17}, 14108 (2014).
\bibitem{Assis}M. Assis, Phys. Rev. E \textbf{98}, 032112 (2018).
\bibitem{XM}P. Di Francesco, E. Guitter ,and S. Mori, Phys. Rev. E \textbf{55}, p.237 (1997). 
\bibitem{FG}P. Di Francesco and E. Guitter, Phys. Rep. \textbf{415}, p. 1–88, (2005).
\bibitem{Kaw_s} T. Kawasaki, in \textit{Proceedings of the First International Meeting of Origami Science and Technology}, (Universita di Padova, Padova, 1989), p. 229-237, J. Justin, in \textit{Proceedings of the First International Meeting of Origami Science and Technology}, (Universita di Padova, Padova, 1989), p. 263-277, T. Kawasaki, British Origami \textbf{(June 1986)} p. 28-30 (1986).
\bibitem{Nis_p}H. Nishimori, Prog. Theor. Phys. \textbf{66}, 1169 (1981).
\bibitem{Nis_t}H. Nishimori, \textit{Statistical Physics of Spin Glasses and Information Processing: An Introduction} (Oxford Univ. Press, Oxford, 2001).
\bibitem{JfH}J. Justin, in \textit{Proceedings of the Second International Meeting of Origami Science and Scientific Origami}, (Seian University of Art and Design, Otsu, 1997), p. 15-29.
\bibitem{MZKST}R. Monasson, R. Zecchina, S. Kirkpatrick, B. Selman, and L. Troyansky, Nature \textbf{400}, 133 (1999). 
\bibitem{WH}M. Weigt and A. K. Hartmann, Phys. Rev. Lett. \textbf{86}, 1658 (2001).
\bibitem{KMRSZ}F. Krzakala, A. Montanari, F. Ricci-Tersenghi, G. Semerjian, and L. Zdeborova, Proc. Natl. Acad. Sci. USA \textbf{104}, 10318 (2007). 
\bibitem{ABDDBSSandM}E. M. Arkin et. al., Comput. Geom. Theory Appl. \textbf{29}, p.23-46 (2004), T. D. Morgan \textit{Map Folding}, PhD Thesis, (MIT, 2012).
\end{references}
\end{document}